\documentstyle[12pt]{article}
  
\catcode`@=11
\def\secteqno{\@addtoreset{equation}{section}%
\def\theequation{\thesection.\arabic{equation}}}
\catcode`@=12
\secteqno
\newcommand{\be}{\begin{equation}}
\newcommand{\ee}{\end{equation}}
\newcommand{\bea}{\begin{eqnarray}}
\newcommand{\eea}{\end{eqnarray}}
\newcommand{\bref}[1]{(\ref{#1})}
\newcommand{\nn}{\nonumber}
\newcommand{\bv}{\mbox{\boldmath $v$}}
\newcommand{\bd}{\mbox{\boldmath $d$}}
\newcommand{\bbeta}{\mbox{\boldmath ${\beta}$}}

\begin{document}
\thispagestyle{empty}
\hfill July 3, 2007

\hfill KEK-TH-1163

\hfill Toho-CP-0784

\vskip 20mm
\begin{center}
{\Large\bf Projective Coordinates and Projective Space Limit}
\vskip 6mm
\medskip

\vskip 10mm
{\large Machiko\ Hatsuda$^{\ast\dagger}$~and~Kiyoshi\ Kamimura$^\star$}

\parskip .15in
{\it $^\ast$Theory Division,\ High Energy Accelerator Research Organization (KEK),\\
\ Tsukuba,\ Ibaraki,\ 305-0801, Japan} \\{\it $^\dagger$Urawa University, Saitama \ 336-0974, Japan}\\
{\small e-mail:\ %\mhref
{mhatsuda@post.kek.jp}} \\
\parskip .35in
{\it 
$~^\star$ 
 Department of Physics, Toho University, Funabashi, 274-8510, Japan}\\
 {\small e-mail:\ %\mhref
 {kamimura@ph.sci.toho-u.ac.jp} }\
 \

\medskip
\end{center}
\vskip 10mm
\begin{abstract}
\end{abstract} 
The ``projective lightcone limit"  has been proposed 
as an alternative holographic  dual of an AdS space.
It is a new type of group contraction for a coset 
G/H
preserving the isometry group G 
but changing H.
In contrast to the usual group contraction,
which changes  G  preserving the spacetime dimension,
it  reduces the dimensions of the spacetime
on which G is realized.
The obtained space is a projective space
on which the isometry 
is realized as a linear  fractional transformation.
We generalize and apply this limiting procedure to the ``Hopf reduction"
and obtain  
$(n$-1)-dimensional complex projective space
from $(2n$-1)-dimensional sphere
preserving SU($n$) symmetry.  

\setcounter{page}{1}
\parskip=7pt
\newpage
%%%%%%%%%%%%%%%%%%%%%%%%%%%%%%%%%%%%%%%%%%%%%%%%%%%%%%%%

\section{Introduction}

In the AdS/CFT correspondence the
global symmetry is one of the most fundamental guiding principles. 
The global SO($D,2$) symmetry is realized 
in terms of not only 
the $(D+1)$-dimensional   AdS space
coordinates but also
 the $D$-dimensional 
conformally flat space coordinates.
In the usual holography this flat $D$-dimensional 
space is located at the boundary of the AdS space \cite{adscft}.
Instead an alternative holography was proposed
\cite{plc1} in which the flat $D$-dimensional space is 
replaced by a lightcone space
obtained by  zero-radius limit of the AdS space
and  
the global symmetry is realized  
by the linear fractional transformations of the projective coordinates
\cite{plcsuper}.
Under the ``projective lightcone limit" 
the $(D+1)$-dimensional AdS metric
reduces into the $D$-dimensional conformally flat metric,
while the AdS metric diverges under the boundary limit 
in the usual holography.
The CFT on the projective lightcone is expected to be newly dual
to the CFT on the usual flat space at the boundary.

The projective lightcone limit is different from the In\"on\"u-Wigner (IW) type
group contraction which does not change the number of generators,
and so the number of coordinates,
but changes the group structure.
The projective lightcone (plc) limit 
changes the number of coordinates preserving the group
 holographically.
The contraction parameter of the plc limit
 is the AdS radius $R$ and 
the limit $R\to 0$ gives a lightcone space.
In the limit the absence of constant scale allows to use projective coordinates
 reducing the number of coordinates.
From the view point of a coset, G/H,  this limit preserves G
but it is a group contraction of H.
The limit is related to  H-covariant quantities
rather than G-covariant quantities;
for a coset element $z\to gzh$ with $g\in $G and $h\in H$
the limiting parameter rescales $z$ from the right
rather than the left.

It was shown that 
the projective lightcone limit of the supersymmetric AdS$_5\times$S$^5$ 
has a possibility to construct the ${N}$=4 SYM theory
on the projective superspace \cite{plcsuper}. 
In order to describe the $N$ extended supersymmetric theories 
SU($N$) internal coordinates are necessary.
The harmonic superspace includes 
the homogeneous coordinates 
for  the SU($N$) symmetry
and harmonic analysis of the
$N$=2, 3 harmonic superspaces has been well 
performed \cite{GIKOS}.
On the other hand the projective superspace \cite{prosusp} includes 
the projective coordinates for  SU($N$)
 and complex analysis is performed.
Originally the projective coordinates are 
used in the K\"{a}hler potential for constructing the non-singular
metric of a manifold and supersymmetric extension 
is obtained by replacing the projective coordinates 
by chiral superfields \cite{Zumino:1979et}.
The $N$=2 projective superspace  is also useful to explore new 
hyperk\"{a}hler metrics and related works are in \cite{Arai:2006gg}.

In this paper we generalize the projective lightcone limit
to a complex projective space limit
where a limiting parameter is introduced besides the AdS radius.
We examine a coset G/H with G=SU($n$) case:
We begin with  a coordinate system  for a
$(2n$-1)-dimensional sphere with the subgroup of the coset 
 H=SU($n$-1), and 
  perform the limit into the $(n$-1)-dimensional complex projective space
where the subgroup becomes H=SU($n$-1)$\otimes$U($1$).  
This limiting procedure from
 $S^{2n-1}$ to $CP^{n-1}$ corresponds
  to the ``Hopf reduction" \cite{Duff:1998us}
which has been studied widely \cite{Nilsson:1984bj}
relating to T-duality in \cite{Duff:9807}, 
to noncompact spaces in \cite{Hori:2002cd}
and to the noncommutative spaces in \cite{fazzycp}.

\par
\vskip 6mm
\section{Generalization of projective lightcone limit}

\subsection{Projective lightcone  limit}

In this section we review the projective lightcone (plc)  limit
clarifying local gauge invariance
and reinterpret it from the group contraction point of view
for a coset.
The plc limit was introduced in
 \cite{plc1} as follows: 
 The $D$-dimensional AdS space is described by a hypersurface in
terms of $(D+1)$-dimensional Minkowski coordinates $x_\mu$ as
\bea
\displaystyle\sum_{\mu=1,\cdots,D,D+1} x_\mu{}^2+R^2=0~~~.\label{adsD1}
\eea
It is rewritten by
projective coordinates $X_i=x_i/x_+$ with 
${i=1,2,\cdots,D-1}$ 
and $U=1/x_+$ where $x_\pm$
are lightcone variables. 
The metric of the D-dimensional AdS space is
\bea
ds^2&=&\displaystyle\sum_{i=1,\cdots,D-1}dx_i^2+dx_+dx_-~=~
 \displaystyle\sum_{i=1,\cdots,D-1}
\displaystyle\frac{dX_i^2}{U^2}+R^2\displaystyle\frac{dU^2}{U^2}
\label{adsmetricR}~~~.
\eea
In the $R\to 0$ limit the hypersurface \bref{adsD1}
becomes  the lightcone space,
and the metric \bref{adsmetricR} reduces into the 
$D-1$-dimensional conformally flat metric with conformal factor $U^{-2}$.
The obtained space is $(D-1)$-dimensional 
lightcone space described by the projective coordinates.  
After the limit 
the coordinate $U$ becomes non-dynamical and 
the dimension of the space is reduced by one.  
$U$ is the dilatation degree of freedom of the $D$-dimensional 
conformal  symmetry.
 
 It was generalized to supersymmetric case in \cite{plcsuper}:
The supersymmetric AdS$_5\times$S$^5$ space  is described by a coset
 GL(4$\mid$4)/(Sp(4)$\otimes$GL(1))$^2$ which is obtained by 
 Wick rotations and 
 introducing gauged degrees of freedom
from a coset
PSU(2,2${\mid}$4)/SO(4,1)$\otimes$SO(5) \cite{Roiban:2000yy}.
After the projective lightcone limit
the coset becomes GL(4$\mid$4)/GL(2$\mid$2)$^2$+ and
the obtained space is  4-dimensional flat space 
with $N=4$ superconformal symmetry which is 
4-dimensional projective lightcone space.

We start with a simple 2-dimensional AdS space.
Its isometry group is SL(2) and it is described by 
parameters of a coset G/H=SL(2)/GL(1).  
For simpler treatment a coset
 GL(2)/GL(1)$^2$ is used by introducing one more coordinate
 with one constraint.
   A  GL(2) matrix is  parametrized as
\bea
z=\left(
\begin{array}{cc}
1&0\\X&{ 1}
\end{array}
\right)
\left(
\begin{array}{cc}
u&0\\0&{ v}
\end{array}
\right)\left(
\begin{array}{cc}
1&Y\\0&{ 1}
\end{array}
\right)
~~~\label{ZZ1}
\eea
with real coordinates $X$, $Y$, $u$ and $v$.
Its inverse is
\bea
z^{-1}=
\left(
\begin{array}{cc}
1&-Y\\0&{1}
\end{array}
\right)
\left(
\begin{array}{cc}
u^{-1}&0\\0&{v}^{-1}
\end{array}
\right)
\left(
\begin{array}{cc}
1&0\\-X&{ 1}
\end{array}
\right)~~~,\label{zinv}
\eea
and the LI one form becomes
\bea
J_A{}^B&=&z^{-1}dz=\left(
\begin{array}{cc}
j_u&j_Y\\j_X&j_v
\end{array}
\right)~\nn\\
&=&\left(
\begin{array}{cc}
\displaystyle\frac{du}{u}-Y\displaystyle\frac{u}{  v}dX
~~~&dY+
\left(\displaystyle\frac{du}{u}
-\displaystyle\frac{dv}{v}\right)Y-\displaystyle\frac{u}{  v}dX Y^2
\\ \\
\displaystyle\frac{u}{  v}dX  
&\displaystyle\frac{dv}{  v}+\displaystyle\frac{u}{  v}dX Y
\end{array}
\right)~~~.\label{zinvdz}
\eea

We choose the basis of  Lie algebra of G and H as follows
\bea
{\cal G}~=~{\rm gl}(2)=\{\tau_{+\rho},~\tau_{-\rho},~\tau_3,~{\bf 1}\}~~,~~
{\cal H}~=~{\rm gl}(1)^2=\{\tau_{+\rho},~{\bf 1}\}\label{HHH}~~~
\eea
where  $\rho$ is a real parameter and
\bea
&\tau_{\pm\rho}=\displaystyle\frac{\tau_+ \pm\rho^2\tau_-}{\rho}=\left(
\begin{array}{cc}
0&1/\rho\\\pm\rho&0
\end{array}
\right)
~~,~~
\tau_\pm=\displaystyle\frac{\tau_1 \pm i\tau_2}{2}\label{hrho}&~~~\\\nn\\
&
\left[\tau_{+\rho},\tau_{-\rho}\right]=-2\tau_3~~,~~
\left[\tau_{\pm \rho},\tau_{3}\right]=-2\tau_{\mp \rho}~~~.\nn
&
\eea
The basis 
$\tau_M=\{\tau_{+\rho},~\tau_{-\rho},~\tau_3,~\tau_{0}={\bf 1}\}$
are normalized as
\bea
\left| (\tau_M)_A{}^B(\tau_N)_C{}^D\Omega^{AC}\Omega_{BD}
\right|=2\delta_{MN}~~~\label{normalization}
\eea
for $\Omega_{AB}=\epsilon_{AB}$. 
The LI one form is decomposed as
\bea
&J_A{}^B=J_M(\tau_M)_A{}^B\label{Jj}&\nn\\
&
J_{\pm\rho}=\displaystyle\frac{1}{2}\left(\rho j_Y\pm 
\displaystyle\frac{j_X}{\rho} \right)~~,~~
J_3=\displaystyle\frac{1}{2}\left(j_u-j_v\right)~~,~~
J_{0}=\displaystyle\frac{1}{2}\left(j_u+j_v\right)&~~~.
\eea
A coset element of G/H of the LI one form is  written as
\bea
\langle J\rangle_A{}^B=J_{-\rho}(\tau_{-\rho})_A{}^B+J_3(\tau_3)_A{}^B
~~~.
\label{rhojrhoj}
\eea
Under the local H-transformation $z \to zh$ with $h \in {\rm H}$
\bea
\langle J\rangle&\to& h^{-1}\langle J\rangle h~~~,
\eea
the bilinear of the coset part current is invariant 
\bea
\langle J\rangle_A{}^B \langle J\rangle_C{}^D \Omega^{AC}\Omega_{BD}
=
\langle J\rangle_A{}^B \langle J\rangle_C{}^D 
\left(h^{-1\ T}\Omega h^{-1}\right)^{AC}
\left(h\Omega h^{T}\right)_{BD}
\eea 
from  $m\Omega m^{T}=(\det m)~\Omega$ for an arbitrary  GL(2)
matrix $m$.
The spacetime metric is 
\bea
ds^2&=&\rho^2 \langle J\rangle_A{}^B \langle J\rangle_C{}^D \Omega^{AC}\Omega_{BD}\nn\\
&=&2\rho^2\left(-J_{-\rho}{}^2 ~+J_3{}^2\right)\nn\\
&=&
\displaystyle\frac{1}{2}
\left\{-\left(\rho^2 j_Y- {j_X} \right)^2
+\rho^2\left(j_u-j_v\right)^2\right\}
\label{metricjjj}
\eea
In the $\rho\to 0$ limit the metric \bref{metricjjj} reduces into 
\bea
ds^2&=&-\displaystyle\frac{1}{2}~
{j_X} ^2~=~\frac{dX^2}{U^2}
\label{metricrho0}~~~
\eea 
with $U=v/u\neq 0$.
This is nothing but the plc metric, 
\bref{adsmetricR} in $R\to 0$ limit.
The global G=GL(2) transformation,  $z\to z'=gz$ with $g\in$ G
is symmetry of the space \bref{metricrho0}  
\bea
 g= \left(\begin{array}{cc}
a&b\\c&d\end{array}\right)~~,~~
X'=\displaystyle\frac{c+dX}{a+bX}~~,~~
U'=\displaystyle\frac{(ad-bc)U}{(a+bX)^2}~~\Rightarrow~~
\frac{d{X'}}{U'}=\frac{d{X}}{U}~~~.\label{AAA}
\eea

In order to trace the local H symmetry relating to 
the local gauge symmetry in the limit
we analyze the system canonically. 
We begin by the Lagrangian for a particle in the coset space 
\bref{metricjjj}
\bea
L&=&\frac{1}{2}\left[-\left\{-\frac{1+\rho^2Y^2}{U}\dot{X}+\rho^2\dot{Y}-
\rho^2\frac{Y\dot{U}}{U}
\right\}^2+
\rho^2\left(-\frac{\dot{U}}{U}-2\frac{Y}{U}\dot{X}\right)^2
\right]
~~~\label{lag}.
\eea 
Only $U$ appears in $L$  resulting GL(2)/GL(1) at this stage.
Conjugate momenta are 
\bea
\left\{\begin{array}{ccl}
p&=&\displaystyle\frac{\partial L}{\partial \dot{X}}~
=~\frac{2\rho}{U}(1+\rho^2Y^2)J_{-\rho}-\frac{4\rho^2 Y}{U}J_3\nn\\\nn\\
\bar{p}&=&\displaystyle\frac{\partial L}{\partial \dot{Y}}~
=~-2\rho^3 J_{-\rho}\nn\\\nn\\
\pi&=&\displaystyle\frac{\partial L}{\partial \dot{U}}~=~
\frac{2\rho^2}{U}(\rho YJ_{-\rho}-J_3)
\end{array}\right.~~.
\eea
The coset part currents are rewritten as 
\bea
J_{-\rho}=-\frac{\bar{p}}{2\rho^3}~~,~~
J_3=-\frac{1}{2\rho^2}\left(Y\bar{p}+U\pi\right)~~~.
\eea
The lack of the kinetic term for $J_{+\rho}$ gives rise to
a primary constraint
\bea
\phi\equiv Up-2U Y\pi+\left(\frac{1}{\rho^2}-Y^2\right)\bar{p}=0~~~.\label{constraint}
\eea
This will be identified with the local H-symmetry generator corresponding to
$\tau_{+\rho}$.
The generators of the local ``right" action are given by
\bea
\phi_{M}=p \delta_{M} X+\bar{p}  \delta_{M}Y+\pi \delta_{M}U
~~,~~z\to ze^{\epsilon^M \tau_M}= z+\delta_M z~~~,
\eea
and they are
\bea
\left\{\begin{array}{ccl}
\phi_{\pm\rho}&=&\rho\left\{
Up\mp 2UY\pi +(\displaystyle\frac{1}{\rho^2}\mp Y^2)\bar{p}
\right\}\\\\
\phi_3&=&-2(Y\bar{p}+U\pi)
\end{array}\right.~~~.\label{phiphi}
\eea
The constraint \bref{constraint} is the local H-transformation generator 
corresponding to  $\tau_{+\rho}$,
$\phi=\phi_{+\rho}/\rho$.

The Hamiltonian is obtained as
\bea
H&=&p\dot{X}+\bar{p}\dot{Y}+\pi\dot{U}-L\nn\\
&=&\displaystyle\frac{1}{2}
\left(\displaystyle\frac{U~p}{1+\rho Y}-\frac{\pi}{\rho U}\right)
\left(\displaystyle\frac{U ~p}{-1+\rho Y}-\frac{\pi}{\rho U}\right)
~~~.
\eea
The  local $\tau_{+\rho}\in{\cal H} $ 
transformation is the gauge symmetry generator
guaranteed by first classness,
$\dot{\phi}=\{\phi,H\}\approx 0$. 
Using this gauge degree of freedom we fix the gauge,
$
Y=0$ with $\{Y,\phi\}\neq 0$,
in such a way that
 the gauge fixed Hamiltonian becomes a simple form
\bea
H_{g.f.}&=&\displaystyle\frac{1}{2}\left(
-U^2p^2+\frac{\pi^2}{\rho^2 U^2}
\right)~~~.
\eea
The gauge fixed Lagrangian becomes
\bea
L_{g.f.}&=&p\dot{X}+\pi\dot{U}-H_{g.f.}
~=~\displaystyle\frac{1}{2}\left(
-\displaystyle\frac{\dot{X}^2}{U^2}
+\rho^2\displaystyle\frac{\dot{U}^2}{U^2}
\right)~~~.
\eea

In the limit $\rho~\to~0$  the 2-dimensional  AdS space \bref{lag}
reduces into the 1-dimensional plc space
\bea
\stackrel
{\rho \to 0}{\longrightarrow} ~~~L_{\rm plc}=-\displaystyle\frac{1}{2}
\displaystyle\frac{\dot{X}^2}{U^2}~~~.\label{metricplc}
\eea
Now $U$ is nondynamical, so
we face to have a new constraint $\pi=0$
originated  to the local $\tau_3$ 
transformation.
The $\phi_{+ \rho}$ transformation constraint
in \bref{phiphi} 
reduce into the 
$\bar{p}= 0$ constraint in $\rho\to 0$ limit.
Using this constraint 
the $\phi_3$ transformation generator 
reduces into $\pi=0$.
The consistency condition requires 
\bea
\dot{\pi}=\left\{\pi,H_{\rm plc}\right\}=Up^2=0~~,~~
H_{\rm plc}=-\frac{1}{2}U^2p^2~~,
\eea
so the invariance of the action $\delta\displaystyle\int  L_{\rm plc}=0$ is given by
\bea
\delta X={\xi}\dot{x}~~,~~\delta U=\xi\dot{U}+\frac{1}{2}\dot{\xi}U~~~.
\eea
The gauge symmetry originated $\tau_3$ transformation 
becomes the 1-dimensional general coordinate transformation
in the plc limit.
The plc system has local gauge invariance.
We regard the local symmetry  generated by
$\bar{p}=0$ and $\pi=0$ 
as those from the stability group of a coset, H, then 
\bea
{\cal G}~=~{\rm gl}(2)=\{\sqrt{2}\tau_{+},~\sqrt{2}\tau_{-},~\tau_3,~{\bf 1}\}~~,~~
{\cal H}~=~{\rm gl}(1)^2+=\{\tau_{3},~{\bf 1},~\sqrt{2}\tau_{+}\}\label{plcHHH}~~~.
\eea
This coset is called ``half coset" which was introduced in \cite{plcsuper};
the subgroup is triangle subgroup where 
diagonal parts are generated by $\tau_3$ and ${\bf 1}$ 
and an upper-right part is generated by $\tau_+$.
The coset is represented only by a lower-left part generated by $\tau_-$.
The factor $\sqrt{2}$ comes from the definition of $\tau_\pm$ in
\bref{hrho} and it is normalized as  \bref{normalization}.
The coset parameter $X$ corresponding to $\tau_{-}$
is a dynamical coordinate of 
 the 1-dimensional space and is transformed under the global 
1-dimensional conformal transformation,  G=GL(2), as \bref{AAA}. 
Although $U$ corresponding to $\tau_3$ is nondynamical in the $\rho \to 0$ limit,
it is indispensable for the G=GL(2) invariance \bref{AAA}.

Let us compare the plc limit with the IW contraction. 
For a Lie group G its  Lie algebra is denoted by ${\cal G}=\left\{T_M\right\}$.
The linear transformation of the generators  $T'_M=V_M{}^NT_N$
does not change the group  if the  transformation is nonsingular, $\det V_M{}^N\neq 0$. 
For the IW contraction the singular transformation is considered in the $\rho \to 0$ 
limit as
$\det V_M{}^N(\rho)=\rho^\nu$ where $\nu$ is the number of the contracted dimension \cite{IW}.
Then  new group G' generated by $\{T'_M\}$ is different from original group G.
On the other hand for the plc limit
the linear transformation is nonsingular even in the $\rho \to 0$ limit
\bea
V_M{}^N=
\left(\begin{array}{ccc}
\frac{1+\rho^2}{2\rho}&\frac{1-\rho^2}{2\rho}&0\\
\frac{1-\rho^2}{2\rho}&\frac{1+\rho^2}{2\rho}&0\\
0&0&1\\
\end{array}\right)~~,~~\det V_M{}^N=1
\eea
where $\{T_M\}=\{\tau_1,i\tau_2,\tau_3\}$ and 
$\{T'_M\}=\{\tau_{+\rho},\tau_{-\rho},\tau_3\}$.
So the plc limit does not change the group G.
However the Lie algebra of  H for a coset G/H 
becomes nilpotent in the $\rho \to 0$ limit.
The coset G/H is a symmetric space for nonzero $\rho$,
but is not so in the $\rho\to 0$ limit
breaking the gauge invariance of the action.
In order to recover the gauge invariance of the action 
the kinetic term for the diagonal part ($\tau_3$ component)
is contracted to ``$0$" and 
the corresponding degree of freedom is gauged.   
As a result the subgroup H is changed to new H'
which is larger than H.
Therefore the number of the  coset parameter for G/H' is  smaller
than the one for G/H.
This subgroup H' is sum of the diagonal part, H'$_0$,  
and the nilpotent part.
Since the number of coset parameters of G/H' is one half of the 
one for G/H'$_{0}$ which is   a symmetric space,
we denote it as the  half coset G/H'$_{0}$+.

\par\vskip 6mm
\subsection{Generalization of projective lightcone  limit}

We generalize the above projective lightcone limit 
to  ``projective space limit"
of a  coset G/H.
A coset element of G/H$\ni z$ is transformed as 
$z~\to~gzh$ with $g\in$ G,~$h\in$ H.

\begin{enumerate}
 \item 
 If a coset element is parametrized as
 \bea
 z=
\left(
\begin{array}{cc}
{  1}&{  0} \\X &{  1} 
\end{array}\right)
\left(
\begin{array}{cc}
u&{  0} \\{  0} &v 
\end{array}\right)
\left(
\begin{array}{cc}
{  1}&Y\\{  0}  &{  1} 
\end{array}\right)
\eea
where $u$ and $v$ are square matrices
and $X$ and $Y$ are rectangular matrices,
then $X$  is  projective coordinate
which is transformed as
\bea
z~&\to&~gz~~,~g=\left(
\begin{array}{cc}
a&b\\c&d
\end{array}
\right)
\nn\\
X~&\to&~(c+dX)(a+bX)^{-1}~~~.\label{231}
\eea
with the following transformation
\bea
u~&\to&~(a+bX)u~~\nn\\
v~&\to&~dv-(c+dX)(a+bX)^{-1}bv\nn\\
Y~&\to&~Y+u^{-1}(a+bX)^{-1}bv~~~.
\eea
The projective coordinate X represents 
the global group G by the linear 
fractional transformation.
\par

\item
There exists a projective space on which the global G symmetry
is represented by the projective
coordinate $X$.
The metric
of the projective space is given by  $ds^2= J_X{}^2$ 
up to normalization, where $J_X$ is the lower-left part
of the  LI one form $z^{-1}dz$ as in the case of
\bref{metricrho0}.
This is obtained by the projective space limit of 
the metric constructed 
in a local H-invariant way
in terms of 
maximal number of coordinates
\bref{metricjjj}.
At first rescale $z$ as
\bea
z~\to~z
\left(
\begin{array}{cc}
1/\sqrt{\rho}&0\\0&\sqrt{\rho}
\end{array}
\right)~~~,
\eea
then the LI one form, $J=z^{-1}dz$, is scaled as
\bea
J~\to~
\left(
\begin{array}{cc}
J_u&\rho J_Y\\
J_X/\rho&
J_v
\end{array}\right)~~~.\label{234}
\eea
Taking $\rho \to 0$ limit in the metric
which is written as bilinear form of the LI currents,
only the $J_X$ component is survived as in \bref{metricrho0}.
\end{enumerate}

\par
\vskip 6mm
\section{Complex projective space limit}

We apply the above procedure
to G=SU($n$) case.
 At first we examine SU($2$) 
as the simplest case.
We present concrete correspondence between SU(2)
coset element and coordinate system of the sphere S$^{3}$.
Then the generalized projective space limit is taken
resulting S$^2$ or CP$^1$. 
Next we examine SU($n$) case.

\subsection{SU(2): S$^3$ to S$^2$}

A 3-dimensional sphere is described by 
three parameters of SU(2).
Instead we use four coordinates and one constraint 
as coset parameters of GL(2)/GL(1) 
which is Wick rotated U(2)/U(1).
A  GL(2) matrix is  parametrized as same as \bref{ZZ1}
\bea
z=\left(
\begin{array}{cc}
1&0\\X&{ 1}
\end{array}
\right)
\left(
\begin{array}{cc}
u&0\\0&{ v}
\end{array}
\right)\left(
\begin{array}{cc}
1&Y\\0&{ 1}
\end{array}
\right)
~~~\label{ZZZ1}
\eea
and $z$ and $z^{-1}dz$ have the same form as \bref{zinv} and \bref{zinvdz}.
Then we go back to U(2) 
by  imposing the unitarity condition on $z$; $z^\dagger z={\bf 1}$.
Its hermite conjugate is given by 
\bea
z^\dagger=
\left(
\begin{array}{cc}
1&0\\Y^\ast&{  1}
\end{array}
\right)
\left(
\begin{array}{cc}
u^{\ast}&0\\0&{  v}^{\ast}
\end{array}
\right)
\left(
\begin{array}{cc}
1&X^\ast\\0&{  1}
\end{array}
\right)~~~.
\eea
The unitarity gives the following relations
\bea
{\mid}u{\mid}^2=\displaystyle\frac{1}{1+{\mid}X{\mid}^2}~~,~~
{\mid}v{\mid}^2=1+{\mid}X{\mid}^2~~,~~
Y=-u^\ast{  v} X^\ast \label{uvXY11}
\eea
with ${\mid}u{\mid}^2=u^\ast u$ and so on.
It leads to $|X|^2={\mid}Y{\mid}^2$, so $Y=0$ gauge can not be chosen
in this case.
 The LI one form satisfies the anti-hermiticity relation,
$\left(z^{-1}dz\right)^\dagger=-z^{-1}dz
$.

The 3-dimensional sphere is parametrized by
SU(2) element $z$ which satisfies
\bea
\displaystyle\sum_{A=0,1}z^\dagger{}_0{}^A z_A{}^0=
 \displaystyle\sum_{A=0,1}z_A{}^0{}^\ast z_B{}^0 \delta^{AB}=1
\eea
 for complex coordinates $z$.
We identify  $z$ with \bref{ZZZ1},
and write down a metric for S$^3$ 
as
\bea
ds^2=\displaystyle\sum_{A,B=0,1}
\left(J_A{}^0\right)^\ast J_B{}^0\delta^{AB} \delta_{00}
~~~\label{metricS3total}.
\eea

The coset element \bref{ZZZ1} is transformed as $z~\to~gz$ with
U(2)$\ni g,~z$ and the LI one forms are manifestly invariant under it.
Under the local U(1) transformation $z\to zh$ with  $h=\left(
\begin{array}{cc}
1&0\\0&e^{i\beta}
\end{array}
\right)$, the LI one form is transformed as
\bea
J_A{}^B~\to~
\left(h^{-1}J h\right)_A{}^{B}
+
\left(
\begin{array}{cc}
0&0\\0&id\beta
\end{array}
\right) ~~~.
\eea
The metric \bref{metricS3total} is invariant under the above U(1) transformation 
from the unitarity condition of $h$,
\bea
\left(h^{-1}{}^\ast\right){}_A{}^C~\left(h^{-1}\right){}_B{}^D~\delta^{AB}=
\delta^{CD}~
~~,~~
\left(h{}^\ast\right){}_0{}^0~\left(h\right){}_0{}^0~\delta_{00}=\delta_{00}~~.
\eea
So the metric of the 3-dimensional sphere \bref{metricS3total}
has both global U(2) symmetry and the local U(1) symmetry.
The first term of the metric \bref{metricS3total}
 becomes
\bea
\left(J_0{}^0\right)^\ast J_0{}^0&=&
\left(\frac{du}{u}-Y\displaystyle\frac{u}{  v}dX\right)^\ast
\left(\frac{du}{u}-Y\displaystyle\frac{u}{  v}dX\right)
\nn\\
&=&\left(d\phi+\displaystyle\frac{i}{2}
\displaystyle\frac{Xd\bar{X}-dX \bar{X}}{1+\mid{X}\mid^2}
\right)^2
\eea
where we use new variables determined from \bref{uvXY11} 
$
u={e^{i\phi}}/{\sqrt{1+{\mid}{X}{\mid}^2}}$.
The second term of the metric \bref{metricS3total} becomes
\bea
\left(J_1{}^0\right)^\ast J_1{}^0&=&
\left(\displaystyle\frac{u}{  v}dX  \right)^\ast 
\left(\displaystyle\frac{u}{  v}dX  \right)=
\displaystyle\frac{{\mid}{dX}{\mid}^2}{(1+{\mid}{X}{\mid}^2)^2}
\label{S2S2}~~~.
\eea
The metric  \bref{S2S2} is nothing but the metric of 
a 2-dimensional sphere. 

Total metric \bref{metricS3total} for a 3-dimensional sphere
is given as 
\bea
ds^2
&=&
\left(d\phi+\displaystyle\frac{i}{2}
\displaystyle\frac{Xd\bar{X}-dX \bar{X}}{1+{\mid}{X}{\mid}^2}
\right)^2
+
\displaystyle\frac{{\mid}{dX}{\mid}^2}{(1+{\mid}{X}{\mid}^2)^2}
\nn\\&=&
\displaystyle\frac{1}{1+{\mid}{\tilde{X}}{\mid}^2}
\left(~d\phi^2
+{\mid}{d\tilde{X}}{\mid}^2
~\right)
-\frac{1}{4}\displaystyle\frac{1}{\left(1+{\mid}\tilde{X}{\mid}^2\right)^2}
d({\mid}{\tilde{X}}{\mid})^2
~~~
\eea
with $\tilde{X}=e^{i\phi}X$.
Changing variables as
$
{\mid}{\tilde{X}}{\mid}^2=r^2$,~$
{\mid}{d\tilde{X}}{\mid}^2=dr^2+r^2d\chi^2
$
it leads to
\bea
ds^2&=&\displaystyle\frac{dr^2}{\left(1+r^2\right)^2}+
\displaystyle\frac{1}{1+r^2}d\phi^2
+\displaystyle\frac{r^2}{1+r^2}d\chi^2~~~.
\eea
Further changing $r=\tan \theta$ leads to
\bea
ds^2=d\theta^2+\cos^2 \theta ~d\phi^2+\sin^2 \theta~d\chi^2~~~
\eea
with $0\leq \theta \leq \pi/2$,~
$-\pi\leq \phi \leq \pi$, $0 \leq \chi\leq \pi$.
This metric represents a 3-dimensional sphere 
which is embedded as
\bea
&x^2+y^2+z^2+w^2=1&\\
&x=\cos \theta\cos \phi,~
y=\cos \theta\sin \phi,~
z=\sin \theta\cos \chi,~
w=\sin \theta\sin \chi&\nn~~~.
\eea
The radius of the sphere $R$ is introduced by replacing
$X$ by $X/R$ and $ds^2$ by $R^2ds^2$ as
\bea
ds^2&=&
R^2\left(d\phi+\displaystyle\frac{i}{2}
\displaystyle\frac{Xd\bar{X}-dX \bar{X}}{R^2+\mid{X}\mid^2}
\right)^2
+
\displaystyle\frac{R^4\mid{dX}\mid^2}{(R^2+\mid{X}\mid^2)^2}
\label{S3S2}
\eea
giving the scalar curvature $4/R^2$. 
In the large radius limit, $R \to \infty$
the curvature becomes zero, and the second term of \bref{S3S2}
reduces into the 2-dimensional flat space while  the first term
becomes one more flat direction with the coordinate $-\infty 
\leq R\phi\leq \infty$.

Now we perform the complex projective space 
limit by following the subsection 2.2.
\begin{enumerate}
  \item As in the equation \bref{231}
  the $X$ is complex projective coordinate which is transformed
under the global U(2) $\ni g$, $g=
\left(\begin{array}{cc}a&b\\c&d\end{array}\right)$ as
\bea
X~\to~X'=\displaystyle\frac{c+dX}{a+bX}~~~.
\eea
\par

\item
As in the equation \bref{234} 
through the rescaling  the coordinates the LI one forms are
scaled as
% as 
% \bea
%z~\to~
%z
%\left(
%\begin{array}{cc}
%1/\sqrt{\rho}&0\\0&\sqrt{\rho}
%\end{array}
%\right)
%\eea
%or equally
%\bea
%X~\to~X~~,~~
%u~\to~\frac{1}{\sqrt{\rho}}u~~,~~v~\to~{\sqrt{\rho}}v~~,~~
%Y~\to~{\rho}Y~~
%\eea
%then 
\bea
J_A{}^B~\to~\left(
\begin{array}{cc}
J_0{}^0&{\rho}J_0{}^1\\
\displaystyle\frac{1}{\rho}J_1{}^0&J_1{}^1
\end{array}\right)~~~.
\eea
%The unitarity condition \bref{uvXY11} is broken 
%by this rescaling with real number $\rho$.
Then the metric in  ${\rho}\to 0$ limit becomes
\bea
ds^2&=&{\rho}^2
R^2\left(d\phi+\displaystyle\frac{i}{2}
\displaystyle\frac{Xd\bar{X}-dX \bar{X}}{R^2+\mid{X}\mid^2}
\right)^2
+
\displaystyle\frac{R^4\mid{dX}\mid^2}{(R^2+\mid{X}\mid^2)^2}
\nn\\
&\stackrel{\rho \to 0}{\longrightarrow}&
\displaystyle\frac{R^4{\mid}{dX}{\mid}^2}{\left(R^2+{\mid}{X}{\mid}^2\right)^2}~~~
\eea
which is the 2-dimensional sphere metric in terms of the complex coordinate.
It is well known that a 2-dimensional sphere is described by  Riemanian
 surface {CP}$^1$;
 the 2-dimensional plane or 1-dimensional complex plane
projected  stereographically of the sphere 
plus a point at infinity.
The resultant coset is U(2)/U(1)$^2$,
since additional constraint $\pi_\phi=0$ corresponds to
additional U(1) in the subgroup.

\end{enumerate}

\par
\vskip 6mm
\subsection{SU(n): S$^{2n-1}$ to CP$^{n-1}$}

Let us consider S$^{2n-1}$ space by taking SU($n$) 
symmetry.
Analogous to the previous section
we use GL($n$)/GL($n$-1) 
instead of
 SU($n$)/SU($n$-1) by Wick rotation and introducing
 gauge coordinates.
The parametrization of GL($n$), $z$, is given by  
as
\bea
z_M{}^A=\left(
\begin{array}{cc}
z_0{}^0&z_0{}^j\\z_i{}^0&z_i{}^j
\end{array}
\right)
=
\left(
\begin{array}{cc}
1&0\\X&{\bf 1}
\end{array}
\right)
\left(
\begin{array}{cc}
u&0\\0&\bv
\end{array}
\right)\left(
\begin{array}{cc}
1&Y\\0&{\bf 1}
\end{array}
\right)
~~~,~~~_{i,j=1,\cdots,n-1}~~~.\label{ZZZ}
\eea
Its inverse is
\bea
z^{-1}=
\left(
\begin{array}{cc}
1&-Y\\0&{\bf 1}
\end{array}
\right)
\left(
\begin{array}{cc}
u^{-1}&0\\0&\bv^{-1}
\end{array}
\right)
\left(
\begin{array}{cc}
1&0\\-X&{\bf 1}
\end{array}
\right)~~~,
\eea
and the Left invariant one form  becomes
\bea
z^{-1}dz&=&
\left(
\begin{array}{cc}
\displaystyle\frac{du}{u}-Y\bv^{-1}dXu
~~~&dY+\displaystyle\frac{du}{u}Y
-Y\bv^{-1}d\bv-Y\bv^{-1}dX uY
\\ \\
\bv^{-1}dX u 
&\bv^{-1}d\bv+\bv^{-1}dXuY
\end{array}
\right)~~~.
\eea
Then we go back to U($n$) by imposing 
the unitarity condition on $z$, 
$z^\dagger z={\bf 1}$
where its hermite conjugate is given by 
\bea
z^\dagger=
\left(
\begin{array}{cc}
1&0\\Y^\dagger&{\bf 1}
\end{array}
\right)
\left(
\begin{array}{cc}
u^{\ast}&0\\0&\bv^{\dagger}
\end{array}
\right)
\left(
\begin{array}{cc}
1&X^\dagger\\0&{\bf 1}
\end{array}
\right)~~~.
\eea
The unitary condition gives the following relations
\bea
{\mid}u{\mid}^2=\displaystyle\frac{1}{1+{\mid}X{\mid}^2}~~&,&~~
Y=-u^\ast X^\dagger \bv
\nn\\
(\bv\bv^\dagger)_i{}^j=\delta_i^j+X_iX^\dagger{}^j=\Upsilon_i{}^j~~&,&~~
\Upsilon^{-1}{}_i{}^j=\delta_i^j-\displaystyle\frac{X_iX^\dagger{}^j}{1+\mid{X}\mid^2}
\label{uvXY}
\eea
satisfying ${\mid}X{\mid}^2={\mid}Y{\mid}^2$ with
 ${\mid}X{\mid}^2=
\displaystyle\sum_{i=1}^{n-1}({X}_i)^\ast X_i$.

A ($2n$-1)-dimensional sphere is parametrized by
SU($n$)/SU($n$-1) parameters as
\bea
\displaystyle\sum_{A=0,1,\cdots,n-1}{z}^\dagger{}_{0}{}^A{} z_{A}{}^0=
\displaystyle\sum_{A=0,1,\cdots,n-1}z_{A}{}^0{}^\ast  z_{B}{}^0\delta^{AB}=
1~~~.
\eea
We identify $z$ with \bref{ZZZ}, and write down a metric of
S$^{2n-1}$ as
\bea
ds^2=\displaystyle\sum_{A,B=0}^{n-1}
\left(J_A{}^0\right)^\ast ~J_B{}^0\delta^{AB}\delta_{00}
~~~.\label{metricSN}
\eea
This is invariant under the local  H transformation:
Under a  H transformation, U($n$-1)$\ni h$, 
$
h=\left(\begin{array}{cc}1&0\\0&\bbeta\end{array}\right)$ with
${\bbeta}^\dagger{\bbeta}={\bf 1}
$
the LI one forms are transformed  as
\bea
J_A{}^B~\to~
\left(h^{-1}Jh\right)_A{}^B
+\left(\begin{array}{cc}0&0\\0&
\bbeta^\dagger d{\bbeta}\end{array}\right)~~~.
\eea
The metric \bref{metricSN} is invariant under $h$ from
  \bea
\left((h^{-1}){}^\ast\right){}_A{}^C~\left(h^{-1}\right){}_B{}^D~\delta^{AB}=
\delta^{CD}~~,~~
\left(h{}^\ast\right){}_0{}^0~\left(h\right)
{}_0{}^0~\delta_{00}=\delta_{00}~~~.
\eea

The first term of the metric \bref{metricSN}
 becomes
\bea
\left(J_0{}^0\right)^\ast ~J_0{}^0&=&
\left[\frac{du}{u}-Y\bv^{-1}dXu\right]^\ast
\left[\frac{du}{u}-Y\bv^{-1}dXu\right]
\nn\\&=&\left(d\phi+A \right)^2\nn\\
A&=&\displaystyle\frac{i}{2}
\displaystyle\frac{\displaystyle\sum_{i=1}^{n-1}
\left(X_id\bar{X}^{i}-dX_i \bar{X}^{i}\right)}{1+{\mid}{X}{\mid}^2}
\eea
where we use $
u={e^{i\phi}}/{\sqrt{1+{\mid}{X}{\mid}^2}}$  
 from \bref{uvXY}.
The rest terms become
\bea
\displaystyle\sum_{i=1}^{n-1}
\left(J_i{}^0\right)^\ast ~J_i{}^0&=&
\displaystyle\sum_{i=1}^{n-1}
\left[\bv^{-1}dX u\right]^{\ast}_i
\left[\bv^{-1}dX u\right]_i\nn\\
&=&\displaystyle\sum_{i,k=1}^{n-1}
\displaystyle\frac{d\bar{X}^{i}}{1+{\mid}{X}{\mid}^2}
\left(
{\bf 1}_{i}{}^k
-\displaystyle\frac{X_i\bar{X}^{ k}}{1+{\mid}{X}{\mid}^2}\right)dX_k
\eea
which is  the Fubini-Study metric for a $(n$-1)-dimensional 
 complex projective space.
The total metric for a $(2n$-1)-dimensional sphere is given by
\bea
ds^2
&=&
\left(d\phi+A
\right)^2
+\displaystyle\sum_{i,k=1}^{n-1}
\displaystyle\frac{d\bar{X}^{i}}{1+{\mid}{X}{\mid}^2}
\left(
{\bf 1}_{i}{}^k
-\displaystyle\frac{X_i\bar{X}^{ k}}{1+{\mid}{X}{\mid}^2}\right)dX_k
\nn\\&=&
\displaystyle\frac{
d\phi^2
+\displaystyle\sum_{i=1}^{n-1}d{\tilde{\bar{X}}}{}^i d\tilde{X}_i
}{1+{\mid}\tilde{X}{\mid}^2}
-\left(\frac{1}{2}\displaystyle\frac{d\displaystyle\sum_{i=1}^{n-1}
{{\tilde{\bar{X}}}}{}^i\tilde{X}_i}{1+{\mid}\tilde{X}{\mid}^2}\right)^2
~~~\label{fubinistudy}
\eea
with $\tilde{X}=e^{i\phi}X$.
Changing variables as
\bea
\mid{\tilde{X}}\mid^2=r^2~~,~~
\mid{d\tilde{X}}\mid^2=dr^2+r^2d\Omega^2_{(2n-3)}
\eea
leads to
\bea
ds^2&=&\displaystyle\frac{dr^2}{(1+r^2)^2}+
\displaystyle\frac{1}{1+r^2}d\phi^2
+\displaystyle\frac{r^2}{1+r^2}d\Omega^2_{(2n-3)}~~~.
\eea
Further rewriting as $r=\tan \theta$
\bea
ds^2=d\theta^2+\cos^2 \theta ~d\phi^2+\sin^2 \theta~d\Omega^2_{(2n-3)}~~~.
\eea
This metric gives constant positive curvature
describing the (2$n$-1)-dimensional sphere.
The radius of the sphere $R$ is inserted back as
\bea
ds^2
&=&
R^2\left(d\phi+A\right)^2
+\displaystyle\sum_{i,k=1}^{n-1}
\displaystyle\frac{R^2d\bar{X}^{i}}{R^2+{\mid}{X}{\mid}^2}
\left(
{\bf 1}_{i}{}^k
-\displaystyle\frac{X_i\bar{X}^{ k}}{R^2+{\mid}{X}{\mid}^2}\right)dX_k
\nn\\\label{2n1}
A&=&
\displaystyle\frac{i}{2}
\displaystyle\frac{\displaystyle\sum_{i=1}^{n-1}
\left(X_id\bar{X}^{i}-dX_i \bar{X}^{i}\right)}{R^2+{\mid}{X}{\mid}^2}
\eea
which reduces into the ($2n$-1)-dimensional flat space metric
in $R \to 0$ limit where the second term in \bref{2n1} 
becomes $(2n$-2)-dimensional flat metric and the first term becomes
 one more coordinate $-\infty \leq R\phi\leq \infty$.

Now let us perform the limiting procedure analogously to the 
previous subsection.
\begin{enumerate}
\item
As in the equation \bref{231}
the $X_i=z_{i}{}^0/z_{0}{}^0$  are  projective coordinates
which are transformed under the global U($n$) transformation as
\bea
X_i~\to~\displaystyle\frac{c_i+\displaystyle\sum_{k=1}^{n-1}\bd_i{}^kX_k}{a+\displaystyle\sum_{j=1}^{n-1}b^jX_j}~~,~~g=
\left(\begin{array}{cc}a&b^j\\c_i&\bd_i{}^j\end{array}\right)\in {\rm U}(n)~~~.
\eea

\item 
As in the equation \bref{234}
through the rescaling the coordinates 
 the LI one forms are rescaled as
\bea
J_A{}^B~\to~\left(
\begin{array}{cc}
J_0{}^0&{\rho}J_0{}^j\\
\displaystyle\frac{1}{\rho}J_i{}^0&J_i{}^j
\end{array}\right)~~~.
\eea
Now let us take the $\rho\to 0$ limit in the metric 
\bea
ds^2
&=&
{\rho}^2R^2\left(d\phi+A\right)^2
+\displaystyle\sum_{i,k=1}^{n-1}
\displaystyle\frac{R^2d\bar{X}^{i}}{R^2+{\mid}{X}{\mid}^2}
\left(
{\bf 1}_{i}{}^k
-\displaystyle\frac{X_i\bar{X}^{ k}}{R^2+{\mid}{X}{\mid}^2}\right)
dX_k 
\nn\\
&\stackrel{\rho\to 0}{\longrightarrow}&\displaystyle\sum_{i,j=1}^{n-1}
\displaystyle\frac{R^2d\bar{X}^i}{R^2+{\mid}{X}{\mid}^2}
\left(
{\bf 1}_{i}{}^j
-\displaystyle\frac{X_i\bar{X}^j}{R^2+{\mid}{X}{\mid}^2}\right)dX_j~~~
\label{fubifubi}~~\\
A&=&\displaystyle\frac{i}{2}
\displaystyle\frac{\displaystyle\sum_{i=1}^{n-1}
\left(X_id\bar{X}^{i}-dX_i \bar{X}^{i}\right)}{R^2+{\mid}{X}{\mid}^2}\nn
\eea
with $\bar{X}^i=X_i{}^\ast$.
Disappearance of the kinetic term for $\phi$ leads to
a new constraint $\pi_\phi=0$ corresponding to additional U(1) 
in the subgroup: G/H with G=U($n$) and H=U($n-1$)$\otimes$U(1).
The obtained metric \bref{fubifubi}  is 
the Fubini-Study metric for the $(n-1)$-dimensional complex projective
space, CP$^{n-1}$.
It is a constant positive curvature space 
but it is not expressed as
the hypersurface in the Euclidean space.
The complex projective space metric is given in terms of 
the  K\"ahler expression
\bea
g_{i\bar{j}}=
\displaystyle\frac{1}{1+{\mid}{X}{\mid}^2}
\left(
{\bf 1}_{i}{}^j
-\displaystyle\frac{X_i\bar{X}^j}{1+{\mid}{X}{\mid}^2}\right)
=\frac{\partial}{\partial \bar{X}^i}
\frac{\partial}{\partial X_j} K
\eea
with the K\"ahler potential 
\bea
K=
\ln \left(
1+{\mid}X{\mid}^2\right)=
-\ln \mid z_0{}^0 \mid^2=-\ln \mid u \mid^2 ~~~,
\eea
from the fact that $\displaystyle\sum_{A=0}^{n-1}{\mid}{z_A{}^0}{\mid}^2=1=
\left(1+\displaystyle\sum_{A=1}^{n-1}{\mid}{X_A{}^0}{\mid}^2\right)
\cdot {\mid}{z_0{}^0}{\mid}^2=\left(1+{\mid}{X}{\mid}^2\right)
\cdot {\mid}{z_0{}^0}{\mid}^2
$.

\end{enumerate}

\par\vskip 6mm
%%%%%%%%%%%%%%%%%%%%%%%%%%%%%%%%%%%%%%%%%%%%%%%%%%%%%%%%%%%%%%%%%%%%%%%%%%%%%%%%%%%%%
\section{Conclusion and discussion}

We have discussed the projective lightcone limit of an AdS space
 with clarifying local symmetries in each step of the limit.
In the plc limit the kinetic term corresponding to the box diagonal element 
is contracted to zero resulting an additional  local gauge symmetry. 
This is regarded as the change of the subgroup H into an upper triangle subgroup.
The coset parameters are reduced into 
 lower triangle matrix elements excluding the box diagonal part, 
 and the number of spacetime coordinate is reduced by one. 
Although the box diagonal element becomes nondynamical,
it is indispensable for realizing the global symmetry G.

We generalize this limit from a sphere to a complex projective space. 
Both spaces have U($n$) symmetry.
A $(2n-1)$-dimensional sphere is described by a coset G/H=U($n$)/U($n-1$),
while a $(n-1)$-dimensional complex projective space is described by
G/H=U($n$)/U($n-1$)$\otimes$U(1).
This projective space limit 
corresponds to the Hopf reduction,
where our method 
is a procedure to relate these spaces
as a kind of group contraction
preserving group symmetries of
projective coordinates manifestly. 
The projective space limit $S^3$ to $S^2$ (CP$^1$) is similar to the gauged
nonlinear sigma model discussed in the subsections 4(C) and 4(D) of the third reference in 
\cite{prosusp}
but different coordinates are used.
Extension to U($n$) case is straightforward for the generalized plc case.
The generalized plc uses a U($n$) matrix as a coordinate, while the gauged   nonlinear sigma model 
uses U($n$) vector. 
Auxiliary degrees of freedom of U($n$) matrix, which are box diagonal parts, 
are essential  to give the Fubini-Study metric \bref{fubinistudy} 
systematically through \bref{uvXY}.
Further applications will be possible to supersymmetric cases, noncompact spaces, noncommutative spaces and T-dual spaces.

%%%%%%%%%%%%%%%%%%%%%%%%%%%%%%%%%%%%%%%%%%%%%%%%%%%%%%%%%%%%%%%%%%%%%%%%%%%%%%%%%%%%%

\section*{Acknowledgments}

We would like to thank Yoji Michishita, 
Shun'ya Mizoguchi, Yu Nakayama, Warren Siegel and Kentaro Yoshida 
for useful discussions.  
M.H. was supported by the Grant-in-Aid for Scientific Research No. 18540287.

%%%%%%%%%%%%%%%%%%%%%%%%%%%%%%%%%%%%%%%%%%%%%%%%%

\end{document}